\documentclass[reprint,prd,showpacs,twoside,superscriptaddress,nofootinbib]{revtex4-1}

\usepackage[T1]{fontenc}
\usepackage[utf8]{inputenc}
\usepackage[brazil,french,english]{babel}
\usepackage{fancyhdr}
  \fancypagestyle{plain}{\fancyhead{}}
\usepackage{amsthm,amsmath,amssymb}
\usepackage{color}
  \definecolor{darkblue}{RGB}{0,0,150}
  \definecolor{RED}{RGB}{255,0,0}
\usepackage[unicode]{hyperref}
 \hypersetup{
   pdfinfo = {
     Author = {Morgan Le Delliou, José P. Mimoso, Filipe C. Mena, Michele Fontanini, Daniel C. Guariento, Elcio Abdalla},
     Title = {Separating expansion and collapse in general fluid models with heat flux}
   },
   colorlinks = true,
   linkcolor = darkblue,
   urlcolor = darkblue,
   citecolor = darkblue
 }
\usepackage{paralist}
\usepackage{tensor}
\usepackage{mathptmx}
\usepackage[mathcal]{euscript}
\usepackage{mathrsfs}
\usepackage{nicefrac}
\DeclareMathAlphabet{\mathbfit}{OML}{cmm}{b}{it}
\usepackage{soul}

\newcommand{\ud}{\ensuremath{\mathrm{d}}}
\newcommand{\lie}{\ensuremath{\mathscr{L}_u}}
\newcommand{\linha}{\ensuremath{^{\prime}}}

\begin{document}

\title{Separating expansion and collapse in general fluid models with heat flux}

\author{Morgan Le Delliou}
\email{Morgan.LeDelliou@uam.es}
\email{delliou@cii.fc.ul.pt}
\affiliation{\foreignlanguage{brazil}{Departamento de Física Matemática, Instituto de Física, Universidade de São Paulo,\\
Caixa Postal 66.318, 05314-970, São Paulo, SP, Brasil}}
\affiliation{\foreignlanguage{brazil}{Centro de Astronomia e Astrofísica da Universidade de Lisboa, Faculdade de Ciências, Ed.~C8, Campo Grande, 1769-016 Lisboa, Portugal}}

\author{José P. Mimoso}
\email{jpmimoso@fc.ul.pt}
\affiliation{\foreignlanguage{brazil}{Centro de Astronomia e Astrofísica da Universidade de Lisboa, Faculdade de Ciências, Ed.~C8, Campo Grande, 1769-016 Lisboa, Portugal}}
\affiliation{\foreignlanguage{brazil}{Departamento de Física da Universidade de Lisboa, Faculdade de Ciências, Ed.~C8, Campo Grande, 1769-016 Lisboa, Portugal}}

\author{Filipe C. Mena}
\email{fmena@math.uminho.pt}
\affiliation{\foreignlanguage{brazil}{Centro de Matemática, Universidade do Minho, Campus de Gualtar, 4710-057 Braga, Portugal}}
\affiliation{Departamento de F\'{i}sica Te\'orica, Instituto de F\'isica, Universidade do Estado do Rio de Janeiro, Rua S\~ao Francisco Xavier, Maracan\~a, 20550-900, Rio de Janeiro, Brazil.}

\author{Michele Fontanini}
\email{fmichele@fma.if.usp.br}
\affiliation{\foreignlanguage{brazil}{Departamento de Física Matemática, Instituto de Física, Universidade de São Paulo,\\
Caixa Postal 66.318, 05314-970, São Paulo, SP, Brasil}}

\author{Daniel C. Guariento}
\email{carrasco@fma.if.usp.br}
\affiliation{\foreignlanguage{brazil}{Departamento de Física Matemática, Instituto de Física, Universidade de São Paulo,\\
Caixa Postal 66.318, 05314-970, São Paulo, SP, Brasil}}

\author{Elcio Abdalla}
\email{eabdalla@fma.if.usp.br}
\affiliation{\foreignlanguage{brazil}{Departamento de Física Matemática, Instituto de Física, Universidade de São Paulo,\\
Caixa Postal 66.318, 05314-970, São Paulo, SP, Brasil}}


\preprint{Version \today}

\date{\today}

\begin{abstract}

In this paper we consider spherically symmetric general fluids with heat flux, motivated by causal thermodynamics, and give the appropriate set of conditions that define separating shells defining the divide between expansion and collapse. To do so we add the new requirement that heat flux and its evolution vanish at the separating surface. We extend previous works with a fully nonlinear analysis in the $1 + 3$ splitting, and present gauge-invariant results. The definition of the separating surface is inspired by the conservation of the Misner-Sharp mass, and is obtained by generalizing the Tolman-Oppenheimer-Volkoff equilibrium and turnaround conditions. We emphasize the nonlocal character of these conditions as found in previous works and discuss connections to the phenomena of spacetime cracking and thermal peeling.

\end{abstract}

\maketitle

\section{Introduction}

The influence of cosmic expansion on local gravitational structures, or the lack thereof, is a long-standing argument of modern gravitation theory, rooted in the works of McVittie and Einstein and Straus \cite{mcvittie-1932,*mcvittie-1933,einstein-1945} (see Ref.~\cite{krasinski} for more historical material). It is related to the general problem of assessing the influence of global physics on local physics \cite{ellis-2002,faraoni-2007} and to the Machian local inertia (see, e.g., Refs.~\cite{sciama-1953,dicke-1961}). Brans-Dicke theory \cite{brans-1961,barrow-1994,mimoso-1995} was indeed produced from these questions.

Einstein-Straus-type models, in particular, have been used to study the influence of cosmic expansion on the solar system \cite{einstein-1945,krasinski}. However, these models have been shown to have a limited scope \cite{mars-2001,mena-2002}. Moreover, they involve matching two metrics, which introduces technical challenges \cite{mars-1993} and tends to hide difficulties at the junction.

The approach we adopt consists of describing the local region embedded in an expanding cosmology using a single metric. A classic example of this viewpoint is provided by the McVittie solution \cite{mcvittie-1933} and the more recent framework of Lasky and Lun \cite{adler-2005,lasky-2006,lasky-2007} using generalized Painlevé-Gullstrand (GPG) coordinates to provide an adequate formalism for this endeavor.  In fact, since the original approach from Lasky and Lun concerns the collapse of compact objects in a vacuum background such as Schwarzschild or Vaidya spacetimes, our approach also applies to this context.

We follow here recent works where a $1 + 3$ spacetime splitting has been used in order to investigate the existence of shells, called \emph{separating shells}, which separate local systems (possibly collapsing) from global expanding regions \cite{mimoso-2010,ledelliou-2011,mimoso-2013}. Previously, only models involving perfect fluids \cite{mimoso-2010} and imperfect fluids with anisotropic stress \cite{mimoso-2013} have been considered. Here, motivated by the Israel-Stewart model \cite{israel-1976,*israel-1979}, we introduce heat flux to the framework and study its role on the existence of separating shells, as well as its impact on other related physical processes.

The intrinsic spatial curvature of the separating shells is shown to be related to the Misner-Sharp mass \cite{misner-1964} and to a function of pressure that generalizes the Tolman-Oppenheimer-Volkoff (TOV) relation of hydrostatic equilibrium \cite{tolman-1939,*oppenheimer-1939a}. These relations are expressed via gauge-invariant conditions, obtained from a novel extension of the treatment originally developed in Refs.~\cite{mimoso-2010,mimoso-2013}. More specifically, we enlarge the validity of those definitions by requiring: 
\begin{inparaenum}[(i)]
\item that there be no matter or heat flux exchange across the shell and
\item that the generalized TOV equation (now including heat flux) be satisfied on that shell, thus ensuring a sort of equilibrium \cite{mimoso-2010,mimoso-2013}.
\end{inparaenum}

In our notation, latin indices run from $0$ to $3$, and $\kappa^2$ is the usual gravitational coupling.

\section{$\mathbf{1+3}$ splitting and gauge-invariant kinematical quantities}\label{sec:splitting-and-gauge}

We present the basic equations in GPG coordinates adapting the works of Lasky and Lun \cite{lasky-2006,lasky-2007} to follow collapse within an underlying overall expansion.

The spherically symmetric line element writes
\begin{equation}\label{eq:dsLaskyLun}
  \ud s^2 = -\alpha^2 \ud t^2 + \frac{\left( \beta \ud t + \ud R \right)^2}{1 + E} + r^2 \ud \Omega^2,
\end{equation}
with $\ud \Omega^2 \equiv \ud \theta^2 + \sin^2 \theta \ud \phi^2$ and $\alpha$, $\beta$, $E$, and $r$ functions of $t$ and $R$. Notice that the areal radius $r$ differs, in principle, from the $R$ coordinate to account for additional degrees of freedom that are required to cope with a general fluid which includes both anisotropic stress and heat flux. The flow of the fluid is characterized by the timelike normalized vector $u_a \equiv -\alpha t_{;a} = -\alpha \tensor*{\delta}{_a^t}$, with $u_a u^a = -1$, which defines a preferred timelike direction, reminiscent of a $3 + 1$ ADM splitting \cite{ellis-1999} with lapse function $N = \alpha$, and the projection operator $h^{ab} \equiv g^{ab} + u^a u^b$ that determines the 3-metric on surfaces normal to the flow. We define the proper-time derivative along the flow of any tensor $\tensor{X}{^a_b}$ as $\tensor{\dot{X}}{^a_b} \equiv u^c \tensor{X}{^a_{b;c}}$. The covariant derivative of the flow defines kinematics, and in general it can be decomposed as $u_{a;b} \equiv -u_b \dot{u}_a + \frac{1}{3} \Theta h_{a b} + \sigma_{a b} + \omega_{a b}$, whose trace gives the expansion $\Theta$, while the symmetric traceless part gives the shear $\sigma_{a b}$. The skew-symmetric vorticity $\omega_{a b}$ vanishes due to spherical symmetry.

The energy-momentum tensor describing the fluid is now
\begin{equation}\label{eq:Tab}
  T_{ab} = \left( \rho + \frac{\Lambda}{8 \pi} \right) u_a u_b + \left( P - \frac{\Lambda}{8 \pi} \right) h_{ab} + 2q_{(a} u_{b)} + \Pi_{ab},
\end{equation}
where $\rho$ is the energy density, $P$ the pressure, $q_a$ the heat flux, $\Pi_{a b}$ the anisotropic stress tensor and $\Lambda$ the cosmological constant. By  definition of heat flux, we have $q^a u_a = 0$, so we can write $q^a = q\, n^a = q (t, R) \sqrt{1 + E} \, \tensor*{\delta}{^a_R}$, where $n^a$ is the spherically-symmetric unit vector in the direction orthogonal to the flow. Moreover, the anisotropic stress tensor satisfies $\Pi^{a b} u_b = 0$ and $\tensor{\Pi}{^a_a} = 0$, and spherical symmetry implies all projected tensors are proportional to the traceless eigenprojector $P_{a b} \equiv h_{a b} - \tensor*{h}{^c_c} n_a n_b$, and so they can be defined by their tangential eigenvalues.

For the Lie derivative of scalars $X$ along $u$, the relations $\lie X = u^a \partial_a X = \dot{X} = \frac{1}{\alpha} \partial_t X - \frac{\beta}{\alpha} \partial_R X$ hold, and from now on we keep the Lie notation and use $X \linha = \partial_R X$. In particular, we note that the generalized expansion function $\mathcal{H} \equiv \frac{\Theta}{3} + \sigma = \frac{\lie r}{r}$, where $\sigma$ is the tangential eigenvalue of the shear.

By using the same projections of Einstein's field equations (EFE) as in Ref.~\cite{mimoso-2013}, we obtain the same forms for the expansion and shear propagations, as well as for the constraint on the Weyl tensor; the latter is induced by the difference in the shear propagation obtained directly from the EFE and Ricci identities. However, the Weyl evolution equation, which comes from the flow-projected Bianchi identity, is modified by the introduction of heat flux, and generalizes the form presented in Ref.~\cite{mimoso-2013} as
\begin{multline}\label{eq:WeylEvol}
  \lie \left( \Xi + \frac{\kappa^2}{2} \Pi \right) = -\frac{\kappa^2}{2} \frac{q}{\mathcal{A}} \lie \sigma\\
  - \frac{\kappa^2}{2} \left\{ \left( \rho + P - 2 \Pi \right) \sigma- q \left[ \frac{\mathcal{A}}{3} - \frac{\sigma}{\mathcal{A}} \left( \Theta - \mathcal{H} \right) \right] \right\}\\
  - \left[2 \left( \Xi + \frac{\kappa^2}{2} \Pi \right) + \left( \Xi - \frac{\kappa^2}{2} \Pi \right) \right] \mathcal{H}.
\end{multline}
Here, $\Xi$ and $\Pi$ are the tangential eigenvalues of the electric part of the Weyl tensor $E_{ab}$ and the anisotropic stress $\Pi_{ab}$, respectively, and $\mathcal{A} \equiv \dot{u}^a n_a = \frac{\alpha \linha}{\alpha}\sqrt{1+E}$ is the projection of the acceleration $\dot{u}^a$ along the unit vector orthogonal to the flow.

The cross-projection of the EFE provides what we call the radial balance constraint
\begin{equation}
  \mathcal{H} \linha + \left( 3 \mathcal{H} - \Theta \right) \frac{r \linha}{r} = \frac{\kappa^2}{2} \frac{q}{\sqrt{1 + E}}, \label{eq:TrjnhEFE:MomentumConstr}\\
\end{equation}
and the orthogonal projections of the Bianchi identities provide the tidal force constraints \cite{lasky-2007}%, which respectively lead to
\begin{equation}
  \frac{\kappa^2}{6} \rho \linha + \frac{\left[ \left( \Xi + \frac{\kappa^2}{2} \Pi \right) r^3 \right] \linha}{r^3} = \frac{\kappa^2}{2} \frac{q}{\sqrt{1 + E}} \mathcal{H}, \label{eq:WeylConstraint}
\end{equation}
whereas the Hamiltonian constraint remains as in Ref.~\cite{mimoso-2013}. The Bianchi identities give the energy density conservation
\begin{equation}\label{eq:nBianchi}
  \lie \rho = - \Theta (\rho + P) - 6 \Pi \, \sigma - q \sqrt{1 + E} \left[ \ln \left( q r^2 \alpha^2 \right) \right] \linha,
\end{equation}
and heat flux conservation, projected along its own direction
\begin{multline}\label{eq:jCons}
  \lie q + 2 \left[ \Theta - \mathcal{H} \right] q = - \left( \rho + P - 2 \Pi \right) \, \mathcal{A}  \\
  + \sqrt{1 + E} \left[ 6 \Pi \frac{r \linha}{r} - \left( P - 2 \Pi \right) \linha \right],
\end{multline}
which plays the role of the heat transport equation in Ref.~\cite{herrera-2009}.

In the presence of heat flux, the radial behavior of the Misner-Sharp mass $M$ changes, following Ref.~\cite{lasky-2007}, to include the corresponding energy exchanges
\begin{equation}\label{MS_mass}
  M \linha = 4 \pi \left( \rho r \linha + \frac{q}{\sqrt{1 + E}} \lie r \right) r^2,
\end{equation}
while its definition remains the same. The areal radius velocity and acceleration are derived in the same fashion as in Ref.~\cite{mimoso-2013}; however, the perfect-fluid-analogous field equations are modified into
\begin{gather}
  r \linha \lie E = 2 \left( 1 + E \right) \left[ \frac{\kappa^2}{2} \frac{q}{\sqrt{1 + E}} r - \left( \lie r \right) \linha - \frac{\beta \linha}{\alpha} r \linha \right],\label{eq:LieE}\\
  \lie M = - \frac{\kappa^2}{2} r^2 \left[ \left( P - 2 \Pi \right) \lie r + q \sqrt{1 + E} \, r \linha \right].\label{eq:LieM}
\end{gather}
Extracting $\nicefrac{\alpha \linha}{\alpha}$ from Eq.~\eqref{eq:jCons}, using the definition of $\mathcal{A}$ and assuming a nonvanishing relativistic inertial mass $\rho + P - 2 \Pi \neq 0$ (note that a negative value, corresponding to a phantomlike fluid, could in principle be considered), we can rewrite the function $\mathrm{gTOV}$ from Ref.~\cite{mimoso-2013} into
\begin{multline}\label{eq:gTOV}
  \mathrm{gTOV} \equiv - {\lie^2 r} = \frac{M}{r^2} + 4 \pi (P - 2 \Pi) r + \frac{\left( 1 + E \right) r \linha}{\rho + P - 2 \Pi} \\
  \times \left[ \left( P - 2 \Pi \right) \linha - 6 \Pi \frac{r \linha}{r} \vphantom{+ \frac{\lie \left( j \sqrt{1 + E} \right)}{\sqrt{1 + E}}} + \frac{\lie q + 2 \left( \Theta - \mathcal{H} \right) q}{\sqrt{1 + E}} \right] - \frac{\Lambda}{3} r,
\end{multline}
which provides a general-relativistic generalization of the Navier-Stokes equation. Setting $\text{gTOV}$ and $\lie r$ to zero reproduces the TOV equilibrium equation. Equation~\eqref{eq:gTOV} shows explicitly the influence of $q$ in the hydrodynamic balance.

\section{Existence of a separating shell}\label{sec:trapped}

We give the conditions for separating shells to exist in the presence of heat flux, discuss their nonlocality connected to the Misner-Sharp mass and analyze the implication on the behavior of temperature profiles in the neighborhood of the shells.

\subsection{Definition with a general fluid}

Now armed with the field equations and the concept of trapped-matter surfaces developed in Refs.~\cite{mimoso-2010,ledelliou-2011,mimoso-2013}, we recognize the local conditions for the existence of a separating surface $r_\star$ to be twofold. first, the heuristic guideline is the conservation of the Misner-Sharp mass, which in the cases previously studied with pressure taken to be nonzero everywhere, corresponded to the conservation of the areal radius along the flow. Following the same heuristic argument, many possibilities open, but the simplest requirement which immediately recovers previous results \cite{mimoso-2010,ledelliou-2011,mimoso-2013} is to impose the additional condition $q_\star = 0$, where the $\star$ denotes a quantity calculated on the separating surface. Here we make this choice, leaving the most general case for future studies. We can see that the requirements of areal radius conservation and vanishing of the heat flux give $(\lie M)_\star = 0$ as from Eq.~\eqref{eq:LieM}.

Second, a relation should come from a generalization of the hydrostatic equilibrium condition which, in parallel to what has been done in Ref.~\cite{mimoso-2013}, generalizes the TOV equation and requires the appropriate $\mathrm{gTOV}_\star = 0$. As can be seen in Eq.~\eqref{eq:gTOV}, the minimal requirement that immediately satisfies the conditions for fluids without heat flux is the vanishing of the flow evolution of the heat flux scalar on the $r_\star$ surface, that is $(\lie q)_\star = 0$.

Recalling that the evolution of the areal radius is linked to the generalized expansion, the above definition of the separating surface, expressed in terms of fully gauge-invariant quantities, is
\\
\begin{subequations}\label{eq:GaugeInvo/LieThetaS3paStar}
\begin{minipage}{.48\linewidth}
  \begin{equation}\label{eq:Hstar0}
    \mathcal{H}_{\star} = 0 ,
  \end{equation}
\end{minipage}
\begin{minipage}{.48\linewidth}
  \begin{equation}\label{eq:lieHstar0}
   ( \lie \mathcal{H})_{\star} = 0,
  \end{equation}
\end{minipage}
\end{subequations}
\\
\begin{subequations}\label{eq:GaugeInvo/LiejStar}
\begin{minipage}{.48\linewidth}
  \begin{equation}\label{eq:lieJstar0}
    q_{\star} = 0,
  \end{equation}
\end{minipage}\hspace{1pt}
\begin{minipage}{.48\linewidth}
  \begin{equation}\label{eq:lie2Jstar0}
    (\lie q)_{\star} = 0 \,.
  \end{equation}
\end{minipage}
\end{subequations}
\vspace{\baselineskip}

Equations~\eqref{eq:Hstar0} and \eqref{eq:lieJstar0} express the conservation of the Misner-Sharp mass or, equivalently, the turning-point condition for the areal radius by analogy with Newtonian dynamics. Equations~\eqref{eq:lieHstar0} and \eqref{eq:lie2Jstar0} encode the hydrostatic balance on the surface. Although local, this definition involves, in the exact manner discussed in Ref.~\cite{mimoso-2013}, nonlocal conditions which are discussed in the next section. Because of the vanishing of heat flux on the $r_\star$ surface by Eqs.~\eqref{eq:GaugeInvo/LiejStar}, the dynamics on the $r_\star$ shell is governed by the same equations as in the anisotropic stress case \cite{mimoso-2013}. However, as we will see ahead, the difference in dynamics of the neighbouring surfaces will have an impact on the existence of the separating shell.

\subsection{Nonlocality with $\mathbfit{q}$}

Introducing heat flux, the expansion reads (here, for convenience, we use $j \equiv \frac{q}{\sqrt{1 + E}}$)
\begin{equation}\label{eq:ThetaInRj}
  \Theta = \frac{\left( \lie r \right) \linha}{r \linha} + 2\frac{\lie r}{r} - \frac{\kappa^2}{2} j \frac{r}{r \linha},
\end{equation}
After choosing a fixed fiducial areal radius $r_{0}$ defined as $r_{0} \equiv r (t, R_{0} (t)) = \text{constant}$, and assuming no shell-crossing at least in the range between $r_0$ and $r_\star$ to maintain the bijection between radius and areal radius, Eq.~\eqref{eq:ThetaInRj} integrates to
\begin{equation}\label{eq:NonLocalTheta}
  \lie r = \frac{1}{r^2} \left[ \int_{r_0}^r \Theta \, r^2 \, \ud r + \frac{\kappa^2}{2} \int_{R_0}^R j r^3 \, \ud R \right] + \frac{1}{r^2} \left( r^2 \lie r \right)_{r_0}.
\end{equation}
The turning-point condition \eqref{eq:Hstar0} at $r_{\star}$ then yields
\begin{equation}\label{eq:NonLocalThetaStar}
  I_0 \equiv - \left( r^2 \lie r \right)_{r_0} - \frac{\kappa^2}{2} \int_{R_0}^{R_{\star}} j r^3 \, \ud R = \int_{r_0}^{r_{\star}} \Theta \, r^2 \, \ud r.
\end{equation}
If the initial parameter $I_0$ vanishes at some interior value $r_0 < r_{\star}$, then so does the right-hand-side integral. This requires the vanishing of the expansion $\Theta$ at some intermediate value $r_0 < \tilde{r} < r_{\star}$, since it has to change signs within the interval. Differentiating equation \eqref{eq:NonLocalTheta} with respect to the flow, we obtain
\begin{align}
  \lie^2 r = & \lie r \left( \Theta - \frac{2}{r} \lie r \right) + \frac{1}{r^2} \left\{ \int_{r_0}^r \lie \Theta \, r^2 \, \ud r \right.\nonumber\\
  & \left. + \lie \left( r^2 \lie r \right)_{r_0} \vphantom{\int_{R_0}^R} \right\}  \nonumber \\
  & + \frac{\kappa^2}{2 \alpha\,r^2} \left[ \int_{R_0}^R \partial_t \left( j r^3 \right) \, \ud R - \left( \beta + \partial_t R_0 \right) j r^3 \right]
   \label{eq:NonLocalTheta_gTOV}\\
  = & - \mathrm{gTOV},\nonumber
\end{align}
which generalizes Eq.~(2.48) of Ref.~\cite{mimoso-2013} and Eq.~(21) of Di Prisco \emph{et al.} \cite{diprisco-1994} and confirms once again the claim that the radial acceleration is nonlocal. From Eq.~\eqref{eq:NonLocalTheta} we realize that this nonlocality is inherent to the radial expansion, and is already present in the energy condition defining $r_{\star}$ [Eqs.~\eqref{eq:Hstar0} and \eqref{eq:lieJstar0}] and in our $\mathrm{gTOV}_{\star} = 0$ condition [Eqs.~\eqref{eq:lieHstar0} and \eqref{eq:lie2Jstar0}] due to the fact that both expressions implicate $M_{\star}$, which is an integral between $0$ and $r_{\star}$ from Eq.~\eqref{MS_mass}.

From the previous Eqs.~\eqref{eq:NonLocalTheta} and \eqref{eq:NonLocalTheta_gTOV}, we see that at the separating shell we have
\begin{align}
  \tilde{I}_0 \equiv & - \frac{\kappa^2}{2 \alpha} \int_{R_0}^{R_{\star}} \partial_t \left( j \, r^3 \right) \, \ud R - \lie \left( r^2 \lie r \right)_{r_0} \nonumber\\
  =& \int_{r_0}^{r_{\star}}  (\lie \Theta) \, r^2 \, \ud r \label{eq:NonLocalTheta_gTOVstar}
\end{align}
so, if $\tilde{I}_0$ vanishes at an interior value $r_0 < r_{\star}$, then so does the right-hand-side integral. This shows that the vanishing of the proper-time derivative of the expansion $\lie \Theta$ occurs at some intermediate value between $r_0$ and $r_{\star}$. When $\tilde{I}_0 = 0$ at the origin, we recover the vanishing of the radial acceleration, i.e.\ $\lie \Theta = 0$ at some $0 < r < r_{\star}$, the result of Di Prisco \emph{et al.} \cite{diprisco-1994}.

\subsection{Transport equations and cracking}

The role of the flows of energy has been analyzed in the literature from the viewpoint of a kinetic theory description of the nonequilibrium processes involved \cite{israel-1976,*israel-1979}.  Originally Eckart and Landau considered a constitutive equation generalizing the Maxwell-Fourier linear relation between heat flow and temperature gradient. However, it is known that the Eckart-Landau law exhibits causality problems arising from the instantaneous propagation of perturbations due to its parabolic nature. To overcome this difficulty, Israel and Stewart \cite{israel-1976,*israel-1979} put forward a heat transport model involving a finite thermal relaxation time (a more general treatment is given in Refs.~\cite{herrera-1997b,sussman-2008,herrera-2009} based on Ref.~\cite{israel-1979}), that leads to
\begin{multline}\label{transp}
  \tau \, \lie q^a + q^a + \tau \, q^b \, \left( \tensor{\sigma}{_b^a} + \frac{\Theta}{3} \, \tensor{h}{_b^a} \right) = K \, h^{ab} \, \left( T_{;b} - T \dot{u}_b \right) \\
  - \frac{1}{2} K \, T^2 \, \left(\frac{\tau}{K T^2}\, u^b \right)_{;b} \, q^a
\end{multline} 
where $K > 0 $ is the heat conduction coefficient, $T$ is the fluid temperature and $\tau$ is the thermal relaxation time.

Imposing conditions \eqref{eq:GaugeInvo/LiejStar} at the separating shell reduces the Israel-Stewart equation to
\begin{equation}
(h^{ab} \, T_{;b})_\star  = (T \dot{u}^a)_\star \; ,
\end{equation}
which allows us to determine the temperature distribution on the shell.

Note that if $\Pi_\star = \Pi\linha_\star = P\linha_\star = 0$ then, from Eq.~\eqref{eq:jCons}, we get $\alpha\linha_\star = 0$ and the geodesic condition $(\dot u^a)_\star = 0$, as expected. In turn, this implies $(h^{ab} \, T_{;b})_\star = 0$, which reflects more clearly the local thermodynamical equilibrium at $r_\star$.

An interesting result, that also illustrates the important role of the flows of energy in connection with separating shells, is provided by the work of Herrera and collaborators \cite{herrera-1997a} who introduced the concept of \emph{thermal peeling} as an effect that contributes to fluid cracking. Their analysis relied on the Maxwell-Fourier law, which is a particular case of Eq.~\eqref{transp} for $\tau = 0$, and reads
\begin{equation}
q^a = K \, h^{ab} \left( T_{;b} - T \dot{u}_b \right). \label{Max-Fourier-law}
\end{equation}
This equation can be projected along $n_a$ to give
\begin{equation}
q = K \, \sqrt{1 + E} \left( T\linha - \frac{\alpha\linha}{\alpha} T \right),
\end{equation}
which, by substituting in \eqref{eq:NonLocalTheta}, leads to 
\begin{multline}\label{eq:Non}
 \lie r = -\frac{1}{r^2} \left[ \int_{r}^{r_\star} \Theta \, r^2 \, \ud r + \frac{\kappa^2K}{2} \left( \vphantom{ \int_{R}^{R_\star} } (T_\star r^3_\star -T r^3)\right.\right.\\
  \left.\left.- \int_{R}^{R_\star} T \frac{(\alpha r^3)^\prime }{\alpha}\right) \, \ud R \right].
\end{multline}
In the proximity of the separating shell, the leading contribution from the second term goes as $(T_\star - T) r^3_\star$; this corroborates the results in Ref.~\cite{herrera-1997a} that sufficiently large negative temperature gradients $T_\star-T<0$ favor cracking and tend to increase the radial velocity of collapse $\lie r>0$ in the interior neighbourhood of $r_\star$. It is important to note though, that cracking shells \cite{herrera-1992} do not have to coincide with separating shells. In fact, comparing the conditions for cracking \cite{herrera-1992,diprisco-1994,mimoso-2013} and Eqs.~\eqref{eq:GaugeInvo/LieThetaS3paStar} and \eqref{eq:GaugeInvo/LiejStar}, we conclude that cracking surfaces also need to satisfy Eq.~\eqref{eq:lie2Jstar0} to be separating shells.

 In this context, as pointed out in Ref.~\cite{herrera-2003}, cracking due to temperature gradients happens only if temperature and heat conductivity are extremely high, otherwise cracking cannot rely on heat flux. Therefore, for small enough temperature and heat conductivity, heat flux can be neglected in most physically relevant cases, and thus, as shown in Ref.~\cite{mimoso-2013}, cracking surfaces are also separating surfaces.

\section{Discussion and Conclusion}

In this work we investigated the existence of separating shells dividing expanding and collapsing regions by using a single metric describing a spherically symmetric inhomogeneous universe with imperfect fluids. The introduction of heat flux has allowed tackling the problem for a wider class of spacetimes than in previous works \cite{mimoso-2013}. Although not fully general, the framework here presented allows treating all cases in which heat flux is segregated on both sides of a separating surface. That is, it makes use of the simplifying conditions \eqref{eq:GaugeInvo/LiejStar}, which can be interpreted as a dynamical constraint on the fluid. The aim of this work has been to explore the main consequences of such an extension, and further discussion will follow in a more general study.

Extending the techniques used previously in Refs.~\cite{adler-2005,lasky-2006,mimoso-2010,ledelliou-2011,mimoso-2013}, we proposed local conditions characterizing the existence of a separating surface. Such a surface, across which heat flux and its evolution along the flow vanish, is defined by locally setting
\begin{inparaenum}[(i)]
\item balance between analogues to total and potential energies, and
\item a generalized TOV hydrostatic equilibrium.
\end{inparaenum}
All this ensures no matter or energy transfer across the separating shell, justifying the present definition to be a valid generalization of the concept of \emph{trapped-matter surfaces} introduced in Ref.~\cite{ledelliou-2011}. The formulation of the intrinsic nonlocality of the shell was also extended, and we constructed the governing equations and separation conditions in terms of gauge invariant scalars as in Ref.~\cite{mimoso-2013}.

We made contact with kinetic theory by using the Israel-Stewart \cite{israel-1976,*israel-1979} relativistic transport equation for dissipative fluids, which has been used before in related problems (see e.g.\ Refs.~\cite{herrera-2010,herrera-2009}). The transport equation applied to our model gives the temperature profile across the separating shell and is consistent with our conditions of local equilibrium. Furthermore, in the limit of vanishing relaxation time, we were able to conclude, as in Ref.~\cite{herrera-1997a}, that temperature gradients can contribute to spacetime cracking. By comparing these conditions with those from Ref.~\cite{mimoso-2013}, we have also established the relationship between cracking and separating shells, and, in the regimes in which they overlap, using Ref.~\cite{herrera-2003} we connected the relationship to a temperature and conductivity scale.

These interesting aspects, especially the role of the temperature gradient in the existence of separating shells, will be studied in more detail in future works.

\begin{acknowledgments}

The work of M.\ Le~D. has been supported by CSIC (JAEDoc072), CICYT (FPA2006-05807) in Spain and FAPESP (2011/24089-5) in Brazil. F.\ C.\ M. thanks CMAT, Univ.\ Minho, for support through FEDER Funds COMPETE, FCT Projects Est-C/MAT/UI0013/2011, PTDC/MAT/108921/2008, CERN/FP/123609/2011 and also Instituto de F\'isica, UERJ, Rio de Janeiro, for hospitality. M.\ Le~D. and J.\ P.\ M. acknowledge the CAAUL's Project No.~PEst-OE/FIS/UI2751/2011. J.\ P.\ M. also wishes to thank FCT for Grants No.~CERN/FP/123615/2011 and CERN/FP/123618/2011. M.\ F. is supported by FAPESP Grant No.~2011/11365-4, D.\ C.\ G. is supported by FAPESP Grant No.~2010/08267-8, and E.\ A. is supported by FAPESP and CNPq.

\end{acknowledgments}

\bibliography{shortnames,referencias}

\begin{thebibliography}{34}%
\makeatletter
\providecommand \@ifxundefined [1]{%
 \@ifx{#1\undefined}
}%
\providecommand \@ifnum [1]{%
 \ifnum #1\expandafter \@firstoftwo
 \else \expandafter \@secondoftwo
 \fi
}%
\providecommand \@ifx [1]{%
 \ifx #1\expandafter \@firstoftwo
 \else \expandafter \@secondoftwo
 \fi
}%
\providecommand \natexlab [1]{#1}%
\providecommand \enquote  [1]{``#1''}%
\providecommand \bibnamefont  [1]{#1}%
\providecommand \bibfnamefont [1]{#1}%
\providecommand \citenamefont [1]{#1}%
\providecommand \href@noop [0]{\@secondoftwo}%
\providecommand \href [0]{\begingroup \@sanitize@url \@href}%
\providecommand \@href[1]{\@@startlink{#1}\@@href}%
\providecommand \@@href[1]{\endgroup#1\@@endlink}%
\providecommand \@sanitize@url [0]{\catcode `\\12\catcode `\$12\catcode
  `\&12\catcode `\#12\catcode `\^12\catcode `\_12\catcode `\%12\relax}%
\providecommand \@@startlink[1]{}%
\providecommand \@@endlink[0]{}%
\providecommand \url  [0]{\begingroup\@sanitize@url \@url }%
\providecommand \@url [1]{\endgroup\@href {#1}{\urlprefix }}%
\providecommand \urlprefix  [0]{URL }%
\providecommand \Eprint [0]{\href }%
\providecommand \doibase [0]{http://dx.doi.org/}%
\providecommand \selectlanguage [0]{\@gobble}%
\providecommand \bibinfo  [0]{\@secondoftwo}%
\providecommand \bibfield  [0]{\@secondoftwo}%
\providecommand \translation [1]{[#1]}%
\providecommand \BibitemOpen [0]{}%
\providecommand \bibitemStop [0]{}%
\providecommand \bibitemNoStop [0]{.\EOS\space}%
\providecommand \EOS [0]{\spacefactor3000\relax}%
\providecommand \BibitemShut  [1]{\csname bibitem#1\endcsname}%
\let\auto@bib@innerbib\@empty
%</preamble>
\bibitem [{\citenamefont {McVittie}(1932)}]{mcvittie-1932}%
  \BibitemOpen
  \bibfield  {author} {\bibinfo {author} {\bibfnamefont {G.~C.}\ \bibnamefont
  {McVittie}},\ }\href {http://adsabs.harvard.edu/abs/1932MNRAS..92..500M}
  {\bibfield  {journal} {\bibinfo  {journal} {Mon. Not. R. Astron. Soc.}\
  }\textbf {\bibinfo {volume} {92}},\ \bibinfo {pages} {500} (\bibinfo {year}
  {1932})}\BibitemShut {NoStop}%
\bibitem [{\citenamefont {McVittie}(1933)}]{mcvittie-1933}%
  \BibitemOpen
  \bibfield  {author} {\bibinfo {author} {\bibfnamefont {G.~C.}\ \bibnamefont
  {McVittie}},\ }\href {http://adsabs.harvard.edu/abs/1933MNRAS..93..325M}
  {\bibfield  {journal} {\bibinfo  {journal} {Mon. Not. R. Astron. Soc.}\
  }\textbf {\bibinfo {volume} {93}},\ \bibinfo {pages} {325} (\bibinfo {year}
  {1933})}\BibitemShut {NoStop}%
\bibitem [{\citenamefont {Einstein}\ and\ \citenamefont
  {Straus}(1945)}]{einstein-1945}%
  \BibitemOpen
  \bibfield  {author} {\bibinfo {author} {\bibfnamefont {A.}~\bibnamefont
  {Einstein}}\ and\ \bibinfo {author} {\bibfnamefont {E.~G.}\ \bibnamefont
  {Straus}},\ }\href {\doibase 10.1103/RevModPhys.17.120} {\bibfield  {journal}
  {\bibinfo  {journal} {Rev. Mod. Phys.}\ }\textbf {\bibinfo {volume} {17}},\
  \bibinfo {pages} {120} (\bibinfo {year} {1945})}\BibitemShut {NoStop}%
\bibitem [{\citenamefont {Krasi\'{n}ski}(1997)}]{krasinski}%
  \BibitemOpen
  \bibfield  {author} {\bibinfo {author} {\bibfnamefont {A.}~\bibnamefont
  {Krasi\'{n}ski}},\ }\href@noop {} {\emph {\bibinfo {title} {Physics in an
  Inhomogeneous Universe}}}\ (\bibinfo  {publisher} {Cambridge University
  Press},\ \bibinfo {address} {Cambridge, England},\ \bibinfo {year}
  {1997})\BibitemShut {NoStop}%
\bibitem [{\citenamefont {Ellis}(2002)}]{ellis-2002}%
  \BibitemOpen
  \bibfield  {author} {\bibinfo {author} {\bibfnamefont {G.~F.~R.}\
  \bibnamefont {Ellis}},\ }\href {\doibase 10.1142/S0217751X02011588}
  {\bibfield  {journal} {\bibinfo  {journal} {Int. J. Mod. Phys. A}\ }\textbf
  {\bibinfo {volume} {17}},\ \bibinfo {pages} {2667} (\bibinfo {year}
  {2002})},\ \Eprint {http://arxiv.org/abs/gr-qc/0102017} {arXiv:gr-qc/0102017}
  \BibitemShut {NoStop}%
\bibitem [{\citenamefont {Faraoni}\ and\ \citenamefont
  {Jacques}(2007)}]{faraoni-2007}%
  \BibitemOpen
  \bibfield  {author} {\bibinfo {author} {\bibfnamefont {V.}~\bibnamefont
  {Faraoni}}\ and\ \bibinfo {author} {\bibfnamefont {A.}~\bibnamefont
  {Jacques}},\ }\href {\doibase 10.1103/PhysRevD.76.063510} {\bibfield
  {journal} {\bibinfo  {journal} {Phys. Rev. D}\ }\textbf {\bibinfo {volume}
  {76}},\ \bibinfo {pages} {063510} (\bibinfo {year} {2007})},\ \Eprint
  {http://arxiv.org/abs/0707.1350} {arXiv:0707.1350 [gr-qc]} \BibitemShut
  {NoStop}%
\bibitem [{\citenamefont {Sciama}(1953)}]{sciama-1953}%
  \BibitemOpen
  \bibfield  {author} {\bibinfo {author} {\bibfnamefont {D.~W.}\ \bibnamefont
  {Sciama}},\ }\href {http://adsabs.harvard.edu/abs/1953MNRAS.113...34S}
  {\bibfield  {journal} {\bibinfo  {journal} {Mon. Not. R. Astron. Soc.}\
  }\textbf {\bibinfo {volume} {113}},\ \bibinfo {pages} {34} (\bibinfo {year}
  {1953})}\BibitemShut {NoStop}%
\bibitem [{\citenamefont {Dicke}(1961)}]{dicke-1961}%
  \BibitemOpen
  \bibfield  {author} {\bibinfo {author} {\bibfnamefont {R.~H.}\ \bibnamefont
  {Dicke}},\ }\href {\doibase 10.1038/192440a0} {\bibfield  {journal} {\bibinfo
   {journal} {Nat.}\ }\textbf {\bibinfo {volume} {192}},\ \bibinfo {pages}
  {440} (\bibinfo {year} {1961})}\BibitemShut {NoStop}%
\bibitem [{\citenamefont {Brans}\ and\ \citenamefont
  {Dicke}(1961)}]{brans-1961}%
  \BibitemOpen
  \bibfield  {author} {\bibinfo {author} {\bibfnamefont {C.}~\bibnamefont
  {Brans}}\ and\ \bibinfo {author} {\bibfnamefont {R.~H.}\ \bibnamefont
  {Dicke}},\ }\href {\doibase 10.1103/PhysRev.124.925} {\bibfield  {journal}
  {\bibinfo  {journal} {Phys. Rev.}\ }\textbf {\bibinfo {volume} {124}},\
  \bibinfo {pages} {925} (\bibinfo {year} {1961})}\BibitemShut {NoStop}%
\bibitem [{\citenamefont {Barrow}\ and\ \citenamefont
  {Mimoso}(1994)}]{barrow-1994}%
  \BibitemOpen
  \bibfield  {author} {\bibinfo {author} {\bibfnamefont {J.~D.}\ \bibnamefont
  {Barrow}}\ and\ \bibinfo {author} {\bibfnamefont {J.~P.}\ \bibnamefont
  {Mimoso}},\ }\href {\doibase 10.1103/PhysRevD.50.3746} {\bibfield  {journal}
  {\bibinfo  {journal} {Phys. Rev. D}\ }\textbf {\bibinfo {volume} {50}},\
  \bibinfo {pages} {3746} (\bibinfo {year} {1994})}\BibitemShut {NoStop}%
\bibitem [{\citenamefont {Mimoso}\ and\ \citenamefont
  {Wands}(1995)}]{mimoso-1995}%
  \BibitemOpen
  \bibfield  {author} {\bibinfo {author} {\bibfnamefont {J.~P.}\ \bibnamefont
  {Mimoso}}\ and\ \bibinfo {author} {\bibfnamefont {D.}~\bibnamefont {Wands}},\
  }\href {\doibase 10.1103/PhysRevD.51.477} {\bibfield  {journal} {\bibinfo
  {journal} {Phys. Rev. D}\ }\textbf {\bibinfo {volume} {51}},\ \bibinfo
  {pages} {477} (\bibinfo {year} {1995})}\BibitemShut {NoStop}%
\bibitem [{\citenamefont {Mars}(2001)}]{mars-2001}%
  \BibitemOpen
  \bibfield  {author} {\bibinfo {author} {\bibfnamefont {M.}~\bibnamefont
  {Mars}},\ }\href {\doibase 10.1088/0264-9381/18/17/317} {\bibfield  {journal}
  {\bibinfo  {journal} {Class. Quantum Gravity}\ }\textbf {\bibinfo {volume}
  {18}},\ \bibinfo {pages} {3645} (\bibinfo {year} {2001})}\BibitemShut
  {NoStop}%
\bibitem [{\citenamefont {C.Mena}\ \emph {et~al.}(2002)\citenamefont {C.Mena},
  \citenamefont {Tavakol},\ and\ \citenamefont {Vera}}]{mena-2002}%
  \BibitemOpen
  \bibfield  {author} {\bibinfo {author} {\bibfnamefont {F.}~\bibnamefont
  {C.Mena}}, \bibinfo {author} {\bibfnamefont {R.}~\bibnamefont {Tavakol}}, \
  and\ \bibinfo {author} {\bibfnamefont {R.}~\bibnamefont {Vera}},\ }\href
  {\doibase 10.1103/PhysRevD.66.044004} {\bibfield  {journal} {\bibinfo
  {journal} {Phys. Rev. D}\ }\textbf {\bibinfo {volume} {66}},\ \bibinfo
  {pages} {044004} (\bibinfo {year} {2002})}\BibitemShut {NoStop}%
\bibitem [{\citenamefont {Mars}\ and\ \citenamefont
  {Senovilla}(1993)}]{mars-1993}%
  \BibitemOpen
  \bibfield  {author} {\bibinfo {author} {\bibfnamefont {M.}~\bibnamefont
  {Mars}}\ and\ \bibinfo {author} {\bibfnamefont {J.~M.~M.}\ \bibnamefont
  {Senovilla}},\ }\href {\doibase 10.1088/0264-9381/10/9/026} {\bibfield
  {journal} {\bibinfo  {journal} {Class. Quantum Gravity}\ }\textbf {\bibinfo
  {volume} {10}},\ \bibinfo {pages} {1865} (\bibinfo {year}
  {1993})}\BibitemShut {NoStop}%
\bibitem [{\citenamefont {Adler}\ \emph {et~al.}(2005)\citenamefont {Adler},
  \citenamefont {Bjorken}, \citenamefont {Chen},\ and\ \citenamefont
  {Liu}}]{adler-2005}%
  \BibitemOpen
  \bibfield  {author} {\bibinfo {author} {\bibfnamefont {R.~J.}\ \bibnamefont
  {Adler}}, \bibinfo {author} {\bibfnamefont {J.~D.}\ \bibnamefont {Bjorken}},
  \bibinfo {author} {\bibfnamefont {P.}~\bibnamefont {Chen}}, \ and\ \bibinfo
  {author} {\bibfnamefont {J.~S.}\ \bibnamefont {Liu}},\ }\href {\doibase
  10.1119/1.2117187} {\bibfield  {journal} {\bibinfo  {journal} {Am. J. Phys.}\
  }\textbf {\bibinfo {volume} {73}},\ \bibinfo {pages} {1148} (\bibinfo {year}
  {2005})},\ \Eprint {http://arxiv.org/abs/gr-qc/0502040} {gr-qc/0502040}
  \BibitemShut {NoStop}%
\bibitem [{\citenamefont {Lasky}\ and\ \citenamefont {Lun}(2006)}]{lasky-2006}%
  \BibitemOpen
  \bibfield  {author} {\bibinfo {author} {\bibfnamefont {P.~D.}\ \bibnamefont
  {Lasky}}\ and\ \bibinfo {author} {\bibfnamefont {A.~W.~C.}\ \bibnamefont
  {Lun}},\ }\href {\doibase 10.1103/PhysRevD.74.084013} {\bibfield  {journal}
  {\bibinfo  {journal} {Phys. Rev. D}\ }\textbf {\bibinfo {volume} {74}},\
  \bibinfo {pages} {084013} (\bibinfo {year} {2006})},\ \Eprint
  {http://arxiv.org/abs/gr-qc/0606055} {arXiv:gr-qc/0606055} \BibitemShut
  {NoStop}%
\bibitem [{\citenamefont {Lasky}\ and\ \citenamefont {Lun}(2007)}]{lasky-2007}%
  \BibitemOpen
  \bibfield  {author} {\bibinfo {author} {\bibfnamefont {P.~D.}\ \bibnamefont
  {Lasky}}\ and\ \bibinfo {author} {\bibfnamefont {A.~W.~C.}\ \bibnamefont
  {Lun}},\ }\href {\doibase 10.1103/PhysRevD.75.024031} {\bibfield  {journal}
  {\bibinfo  {journal} {Phys. Rev. D}\ }\textbf {\bibinfo {volume} {75}},\
  \bibinfo {pages} {024031} (\bibinfo {year} {2007})},\ \Eprint
  {http://arxiv.org/abs/gr-qc/0612007} {arXiv:gr-qc/0612007} \BibitemShut
  {NoStop}%
\bibitem [{\citenamefont {Mimoso}\ \emph {et~al.}(2010)\citenamefont {Mimoso},
  \citenamefont {{Le Delliou}},\ and\ \citenamefont {Mena}}]{mimoso-2010}%
  \BibitemOpen
  \bibfield  {author} {\bibinfo {author} {\bibfnamefont {J.~P.}\ \bibnamefont
  {Mimoso}}, \bibinfo {author} {\bibfnamefont {M.}~\bibnamefont {{Le
  Delliou}}}, \ and\ \bibinfo {author} {\bibfnamefont {F.~C.}\ \bibnamefont
  {Mena}},\ }\href {\doibase 10.1103/PhysRevD.81.123514} {\bibfield  {journal}
  {\bibinfo  {journal} {Phys. Rev. D}\ }\textbf {\bibinfo {volume} {81}},\
  \bibinfo {pages} {123514} (\bibinfo {year} {2010})},\ \Eprint
  {http://arxiv.org/abs/0910.5755} {arXiv:0910.5755 [gr-qc]} \BibitemShut
  {NoStop}%
\bibitem [{\citenamefont {{Le Delliou}}\ \emph {et~al.}(2011)\citenamefont {{Le
  Delliou}}, \citenamefont {Mena},\ and\ \citenamefont
  {Mimoso}}]{ledelliou-2011}%
  \BibitemOpen
  \bibfield  {author} {\bibinfo {author} {\bibfnamefont {M.}~\bibnamefont {{Le
  Delliou}}}, \bibinfo {author} {\bibfnamefont {F.~C.}\ \bibnamefont {Mena}}, \
  and\ \bibinfo {author} {\bibfnamefont {J.~P.}\ \bibnamefont {Mimoso}},\
  }\href {\doibase 10.1103/PhysRevD.83.103528} {\bibfield  {journal} {\bibinfo
  {journal} {Phys. Rev. D}\ }\textbf {\bibinfo {volume} {83}},\ \bibinfo
  {pages} {103528} (\bibinfo {year} {2011})},\ \Eprint
  {http://arxiv.org/abs/1103.0976} {arXiv:1103.0976 [gr-qc]} \BibitemShut
  {NoStop}%
\bibitem [{\citenamefont {Mimoso}\ \emph {et~al.}(2013)\citenamefont {Mimoso},
  \citenamefont {{Le Delliou}},\ and\ \citenamefont {Mena}}]{mimoso-2013}%
  \BibitemOpen
  \bibfield  {author} {\bibinfo {author} {\bibfnamefont {J.~P.}\ \bibnamefont
  {Mimoso}}, \bibinfo {author} {\bibfnamefont {M.}~\bibnamefont {{Le
  Delliou}}}, \ and\ \bibinfo {author} {\bibfnamefont {F.~C.}\ \bibnamefont
  {Mena}},\ }\href {\doibase 10.1103/PhysRevD.88.043501} {\bibfield  {journal}
  {\bibinfo  {journal} {Phys. Rev. D}\ }\textbf {\bibinfo {volume} {88}},\
  \bibinfo {pages} {043501} (\bibinfo {year} {2013})},\ \Eprint
  {http://arxiv.org/abs/1302.6186} {arXiv:1302.6186 [gr-qc]} \BibitemShut
  {NoStop}%
\bibitem [{\citenamefont {Israel}\ and\ \citenamefont
  {Stewart}(1976)}]{israel-1976}%
  \BibitemOpen
  \bibfield  {author} {\bibinfo {author} {\bibfnamefont {W.}~\bibnamefont
  {Israel}}\ and\ \bibinfo {author} {\bibfnamefont {J.~M.}\ \bibnamefont
  {Stewart}},\ }\href {\doibase 10.1016/0375-9601(76)90075-X} {\bibfield
  {journal} {\bibinfo  {journal} {Phys. Lett. A}\ }\textbf {\bibinfo {volume}
  {58}},\ \bibinfo {pages} {213} (\bibinfo {year} {1976})}\BibitemShut
  {NoStop}%
\bibitem [{\citenamefont {Israel}\ and\ \citenamefont
  {Stewart}(1979)}]{israel-1979}%
  \BibitemOpen
  \bibfield  {author} {\bibinfo {author} {\bibfnamefont {W.}~\bibnamefont
  {Israel}}\ and\ \bibinfo {author} {\bibfnamefont {J.~M.}\ \bibnamefont
  {Stewart}},\ }\href {\doibase 10.1016/0003-4916(79)90130-1} {\bibfield
  {journal} {\bibinfo  {journal} {Ann. Phys.}\ }\textbf {\bibinfo {volume}
  {118}},\ \bibinfo {pages} {341} (\bibinfo {year} {1979})}\BibitemShut
  {NoStop}%
\bibitem [{\citenamefont {Misner}\ and\ \citenamefont
  {Sharp}(1964)}]{misner-1964}%
  \BibitemOpen
  \bibfield  {author} {\bibinfo {author} {\bibfnamefont {C.~W.}\ \bibnamefont
  {Misner}}\ and\ \bibinfo {author} {\bibfnamefont {D.~H.}\ \bibnamefont
  {Sharp}},\ }\href {\doibase 10.1103/PhysRev.136.B571} {\bibfield  {journal}
  {\bibinfo  {journal} {Phys. Rev.}\ }\textbf {\bibinfo {volume} {136}},\
  \bibinfo {pages} {B571} (\bibinfo {year} {1964})}\BibitemShut {NoStop}%
\bibitem [{\citenamefont {Tolman}(1939)}]{tolman-1939}%
  \BibitemOpen
  \bibfield  {author} {\bibinfo {author} {\bibfnamefont {R.~C.}\ \bibnamefont
  {Tolman}},\ }\href {\doibase 10.1103/PhysRev.55.364} {\bibfield  {journal}
  {\bibinfo  {journal} {Phys. Rev.}\ }\textbf {\bibinfo {volume} {55}},\
  \bibinfo {pages} {364} (\bibinfo {year} {1939})}\BibitemShut {NoStop}%
\bibitem [{\citenamefont {Oppenheimer}\ and\ \citenamefont
  {Volkoff}(1939)}]{oppenheimer-1939a}%
  \BibitemOpen
  \bibfield  {author} {\bibinfo {author} {\bibfnamefont {J.~R.}\ \bibnamefont
  {Oppenheimer}}\ and\ \bibinfo {author} {\bibfnamefont {G.~M.}\ \bibnamefont
  {Volkoff}},\ }\href {\doibase 10.1103/PhysRev.55.374} {\bibfield  {journal}
  {\bibinfo  {journal} {Phys. Rev.}\ }\textbf {\bibinfo {volume} {55}},\
  \bibinfo {pages} {374} (\bibinfo {year} {1939})}\BibitemShut {NoStop}%
\bibitem [{\citenamefont {Ellis}\ and\ \citenamefont {van
  Elst}(1999)}]{ellis-1999}%
  \BibitemOpen
  \bibfield  {author} {\bibinfo {author} {\bibfnamefont {G.~F.~R.}\
  \bibnamefont {Ellis}}\ and\ \bibinfo {author} {\bibfnamefont
  {H.}~\bibnamefont {van Elst}},\ }in\ \href@noop {} {\emph {\bibinfo
  {booktitle} {Proceedings of the NATO Advanced Study Institute on Theoretical
  and Observational Cosmology, Carg\`{e}se, France, August 17-29, 1998}}},\
  \bibinfo {series} {NATO science series. Series C, Mathematical and physical
  sciences}, Vol.\ \bibinfo {volume} {541},\ \bibinfo {editor} {edited by\
  \bibinfo {editor} {\bibfnamefont {M.}~\bibnamefont {Lachi\`{e}ze-Rey}}}\
  (\bibinfo  {publisher} {Kluwer Academic},\ \bibinfo {address} {Boston},\
  \bibinfo {year} {1999})\ pp.\ \bibinfo {pages} {1--116},\ \Eprint
  {http://arxiv.org/abs/gr-qc/9812046} {arXiv:gr-qc/9812046} \BibitemShut
  {NoStop}%
\bibitem [{\citenamefont {Herrera}\ \emph {et~al.}(2009)\citenamefont
  {Herrera}, \citenamefont {Prisco}, \citenamefont {Fuenmayor},\ and\
  \citenamefont {Troconis}}]{herrera-2009}%
  \BibitemOpen
  \bibfield  {author} {\bibinfo {author} {\bibfnamefont {L.}~\bibnamefont
  {Herrera}}, \bibinfo {author} {\bibfnamefont {A.~D.}\ \bibnamefont {Prisco}},
  \bibinfo {author} {\bibfnamefont {E.}~\bibnamefont {Fuenmayor}}, \ and\
  \bibinfo {author} {\bibfnamefont {O.}~\bibnamefont {Troconis}},\ }\href
  {\doibase 10.1142/S0218271809014285} {\bibfield  {journal} {\bibinfo
  {journal} {Int. J. Mod. Phys. D}\ }\textbf {\bibinfo {volume} {18}},\
  \bibinfo {pages} {129} (\bibinfo {year} {2009})}\BibitemShut {NoStop}%
\bibitem [{\citenamefont {{Di Prisco}}\ \emph {et~al.}(1994)\citenamefont {{Di
  Prisco}}, \citenamefont {Fuenmayor}, \citenamefont {Herrera},\ and\
  \citenamefont {Varela}}]{diprisco-1994}%
  \BibitemOpen
  \bibfield  {author} {\bibinfo {author} {\bibfnamefont {A.}~\bibnamefont {{Di
  Prisco}}}, \bibinfo {author} {\bibfnamefont {E.}~\bibnamefont {Fuenmayor}},
  \bibinfo {author} {\bibfnamefont {L.}~\bibnamefont {Herrera}}, \ and\
  \bibinfo {author} {\bibfnamefont {V.}~\bibnamefont {Varela}},\ }\href
  {\doibase 10.1016/0375-9601(94)90420-0} {\bibfield  {journal} {\bibinfo
  {journal} {Phys. Lett. A}\ }\textbf {\bibinfo {volume} {195}},\ \bibinfo
  {pages} {23} (\bibinfo {year} {1994})}\BibitemShut {NoStop}%
\bibitem [{\citenamefont {Herrera}\ \emph {et~al.}(1997)\citenamefont
  {Herrera}, \citenamefont {Di~Prisco}, \citenamefont {Hernández-Pastora},
  \citenamefont {Martín},\ and\ \citenamefont {Martínez}}]{herrera-1997b}%
  \BibitemOpen
  \bibfield  {author} {\bibinfo {author} {\bibfnamefont {L.}~\bibnamefont
  {Herrera}}, \bibinfo {author} {\bibfnamefont {A.}~\bibnamefont {Di~Prisco}},
  \bibinfo {author} {\bibfnamefont {J.~L.}\ \bibnamefont {Hernández-Pastora}},
  \bibinfo {author} {\bibfnamefont {J.}~\bibnamefont {Martín}}, \ and\
  \bibinfo {author} {\bibfnamefont {J.}~\bibnamefont {Martínez}},\ }\href
  {\doibase 10.1088/0264-9381/14/8/022} {\bibfield  {journal} {\bibinfo
  {journal} {Class. Quantum Gravity}\ }\textbf {\bibinfo {volume} {14}},\
  \bibinfo {pages} {2239} (\bibinfo {year} {1997})},\ \Eprint
  {http://arxiv.org/abs/gr-qc/9704022} {arXiv:gr-qc/9704022} \BibitemShut
  {NoStop}%
\bibitem [{\citenamefont {Sussman}(2008)}]{sussman-2008}%
  \BibitemOpen
  \bibfield  {author} {\bibinfo {author} {\bibfnamefont {R.~A.}\ \bibnamefont
  {Sussman}},\ }in\ \href@noop {} {\emph {\bibinfo {booktitle} {The Casimir
  effect and Cosmology: A volume in honour of Professor {Iver H. Brevik} on the
  occasion of his 70th birthday}}},\ \bibinfo {editor} {edited by\ \bibinfo
  {editor} {\bibfnamefont {S.~D.}\ \bibnamefont {Odintsov}}, \bibinfo {editor}
  {\bibfnamefont {E.}~\bibnamefont {Elizalde}}, \ and\ \bibinfo {editor}
  {\bibfnamefont {O.~G.}\ \bibnamefont {Gorbunova}}}\ (\bibinfo  {publisher}
  {TSPU Publishing Company},\ \bibinfo {address} {Tomsk},\ \bibinfo {year}
  {2008})\ pp.\ \bibinfo {pages} {125--141},\ \Eprint
  {http://arxiv.org/abs/0812.4430} {arXiv:0812.4430 [gr-qc]} \BibitemShut
  {NoStop}%
\bibitem [{\citenamefont {Herrera}\ and\ \citenamefont
  {Di~Prisco}(1997)}]{herrera-1997a}%
  \BibitemOpen
  \bibfield  {author} {\bibinfo {author} {\bibfnamefont {L.}~\bibnamefont
  {Herrera}}\ and\ \bibinfo {author} {\bibfnamefont {A.}~\bibnamefont
  {Di~Prisco}},\ }\href {\doibase 10.1103/PhysRevD.55.2044} {\bibfield
  {journal} {\bibinfo  {journal} {Phys. Rev. D}\ }\textbf {\bibinfo {volume}
  {55}},\ \bibinfo {pages} {2044} (\bibinfo {year} {1997})}\BibitemShut
  {NoStop}%
\bibitem [{\citenamefont {Herrera}(1992)}]{herrera-1992}%
  \BibitemOpen
  \bibfield  {author} {\bibinfo {author} {\bibfnamefont {L.}~\bibnamefont
  {Herrera}},\ }\href {\doibase 10.1016/0375-9601(92)90036-L} {\bibfield
  {journal} {\bibinfo  {journal} {Phys. Lett. A}\ }\textbf {\bibinfo {volume}
  {165}},\ \bibinfo {pages} {206} (\bibinfo {year} {1992})}\BibitemShut
  {NoStop}%
\bibitem [{\citenamefont {Herrera}\ and\ \citenamefont
  {Santos}(2003)}]{herrera-2003}%
  \BibitemOpen
  \bibfield  {author} {\bibinfo {author} {\bibfnamefont {L.}~\bibnamefont
  {Herrera}}\ and\ \bibinfo {author} {\bibfnamefont {N.~O.}\ \bibnamefont
  {Santos}},\ }\href {\doibase 10.1046/j.1365-8711.2003.06764.x} {\bibfield
  {journal} {\bibinfo  {journal} {Mon. Not. R. Astron. Soc.}\ }\textbf
  {\bibinfo {volume} {343}},\ \bibinfo {pages} {1207} (\bibinfo {year}
  {2003})}\BibitemShut {NoStop}%
\bibitem [{\citenamefont {Herrera}\ \emph {et~al.}(2010)\citenamefont
  {Herrera}, \citenamefont {{Di Prisco}},\ and\ \citenamefont
  {Ospino}}]{herrera-2010}%
  \BibitemOpen
  \bibfield  {author} {\bibinfo {author} {\bibfnamefont {L.}~\bibnamefont
  {Herrera}}, \bibinfo {author} {\bibfnamefont {A.}~\bibnamefont {{Di
  Prisco}}}, \ and\ \bibinfo {author} {\bibfnamefont {J.}~\bibnamefont
  {Ospino}},\ }\href {\doibase 10.1007/s10714-010-0931-6} {\bibfield  {journal}
  {\bibinfo  {journal} {Gen. Relativ. Gravit.}\ }\textbf {\bibinfo {volume}
  {42}},\ \bibinfo {pages} {1585} (\bibinfo {year} {2010})},\ \Eprint
  {http://arxiv.org/abs/1001.3020} {arXiv:1001.3020 [gr-qc]} \BibitemShut
  {NoStop}%
\end{thebibliography}%

\end{document}